\documentclass[twocolumn,10pt]{IEEEtran}
\usepackage{ifpdf, flushend,subfigure}

\ifCLASSINFOpdf
  \usepackage[pdftex]{graphicx}
  \graphicspath{{../pdf/}{../jpeg/}}
  \DeclareGraphicsExtensions{.pdf,.jpeg,.png}
\else
  \usepackage[dvips]{graphicx}
  \graphicspath{{../eps/}}
  \DeclareGraphicsExtensions{.eps}
\fi
\usepackage{graphicx}
\usepackage{epstopdf}
\graphicspath{{../}}
\usepackage{multirow}
\usepackage{float}
\usepackage[cmex10]{amsmath}
\usepackage {amssymb}
\usepackage{array}
\usepackage{mdwmath}
\usepackage{mdwtab}
\usepackage{eqparbox}
\usepackage{nomencl}
\usepackage{cite}
\usepackage{algorithm}
\usepackage{algorithmic}
\usepackage{makeidx}
\usepackage{ifthen}
\usepackage{subfigure}
\usepackage{caption}
\usepackage{pdflscape}
\usepackage{xcolor}
\newcommand{\beq}{\begin{equation}}
\newcommand{\eeq}{\end{equation}}
\newcommand{\beqn}{\begin{eqnarray}}
\newcommand{\eeqn}{\end{eqnarray}}
\newcommand{\beqno}{\begin{eqnarray*}}
\newcommand{\eeqno}{\end{eqnarray*}}
\newcommand{\bma}{\begin{displaymath}}
\newcommand{\ema}{\end{displaymath}}
\newcommand{\bnu}{\begin{enumerate}}
\newcommand{\enu}{\end{enumerate}}
\newcommand{\bce}{\begin{center}}
\newcommand{\ece}{\end{center}}
\newcommand{\btb}{\begin{tabular}}
\newcommand{\etb}{\end{tabular}}

\usepackage[top=0.75in,bottom=2.8cm,left=0.65in,right=0.65in]{geometry}
\usepackage{amsmath}
\usepackage[bookmarks=false]{hyperref}
\begin{document}
\title{Coordinated Deliverable Energy Flexibility from EV Aggregators in Distribution Networks}

\author{\IEEEauthorblockN{Arash~Baharvandi, ~Duong~Tung~Nguyen,~\IEEEmembership{Member,~IEEE}}


 }


\maketitle

\begin{abstract}
This paper presents a coordinated framework to optimize electric vehicle (EV) charging considering grid constraints and system uncertainties. The proposed framework consists of two optimization models. In particular, the distribution system operator (DSO) solves the first model to optimize the amount of deliverable energy flexibility that can be obtained from EV aggregators. To address the uncertainties of loads and solar energy generation,  a hybrid robust/stochastic approach is employed, enabling the transformation of uncertainty-related constraints into a set of equivalent deterministic constraints. 
Once the DSO has computed the optimal energy flexibility, each aggregator utilizes the second optimization model to optimize the charging schedule for its respective fleet of EVs.
Numerical simulations are performed  on a modified IEEE 33-bus distribution network to illustrate the efficiency of the proposed framework.

\end{abstract}

\vspace{-0.5cm}

\section{Introduction}



The increasing popularity of electric vehicles (EVs) can be attributed to their environmental benefits and cost-effectiveness. However, integrating EVs effectively into power grids poses a fundamental challenge. Notably, the deliverable energy flexibility from EV charging can enhance grid operations by transferring it to upstream networks 
 \cite{masood19}. 
Additionally, EVs can impact the grid's voltage, stability, and losses \cite{hu2016electric}. For instance, 
 bi-directional charging and vehicle-to-grid (V2G) technologies enable EVs to support the grid by participating in frequency regulation and voltage correction   \cite{amamra2019vehicle}. 
 Indeed, the integration of EV charging into grid operations has been extensively explored in existing literature.

Various  studies have addressed coordinated EV charging to support the grid in different ways
\cite{kaur2018coordinated,mazumder2020ev,ladhiya2024design,singh2019time,nizami2020coordinated}. In \cite{kaur2018coordinated}, a power management approach,  considering bidirectional V2G capabilities, is proposed to minimize frequency deviation while optimizing EV charging  and EV owners' revenue. 
However, incorporating EVs into the grid may not always yield feasible solutions, as discussed in  \cite{mazumder2020ev}, where the focus is on minimizing EV charging costs, employing both slow and fast chargers.
Charging multiple high-power EV batteries simultaneously can cause power quality problems like overloads, power gaps, and voltage drops. Reference \cite{ladhiya2024design} introduces a coordinated control method for EV charging stations to address these issues.
In \cite{singh2019time}, conservation voltage regulation (CVR) is explored, along with the incorporation of EVs into the distribution network. A volt/var optimization (VVO) method is defined to examine the combined effects of EVs and CVR on the network.
In \cite{nizami2020coordinated}, a coordinated model is introduced for EVs in low voltage networks, aiming to mitigate voltage deviations and overloaded equipment. 


Numerous existing studies have proposed deterministic EV scheduling models without considering any system uncertainty 
\cite{liu2020optimal,zhang2018optimal,elghanam2024optimization,liu2018two}. For example, an optimal scheduling approach is presented  in \cite{liu2020optimal}, incorporating time-of-use (TOU) electricity pricing. It considers uncontrolled charging for fast EV charging, while minimizing costs by limiting the number of chargers to meet demand. 
Similarly, reference   \cite{zhang2018optimal} formulates an optimal pricing problem to reduce the number of EVs left uncharged and abandoned at charging stations, aiming to minimize the service dropping rate.
In \cite{elghanam2024optimization}, the authors review literature on deterministic optimization techniques for EV charging coordination, focusing on time-based scheduling, spatial coordination, and spatio-temporal strategies. 
Reference \cite{liu2018two} applies a two-stage scheduling with transactive control to manage day-ahead electricity purchases and real-time EV charging, with the objective of minimizing charging costs.

In practice, it is crucial to account for various system uncertainties when optimizing EV charging models  \cite{wang2019interval}.
There is a rich literature \cite{fallah2020charge,han2020optimal,barhagh2023optimal,shi2022day,huang2016robust}  that has taken uncertainties into account by incorporating stochastic programming (SP) and robust optimization (RO) methods in problems involving EV integration. In \cite{fallah2020charge}, a multi-stage stochastic programming approach is utilized to minimize the energy cost  over a limited horizon, considering uncertain future demand. 
Similarly, a multi-stage stochastic optimization technique  is employed  in \cite{han2020optimal} to enable energy arbitrage in the energy market, 
focusing on using the collective power of aggregated EVs to offer ancillary services and maximize profits in the future market.
 Reference \cite{barhagh2023optimal} introduces an optimization framework that incorporates uncertainty, utilizing robust optimization and scenario analysis methods to determine the optimal sizing and placement of electric vehicle charging stations (EVCSs). 
Reference \cite{shi2022day} utilize RO to explore a day-ahead EV scheduling approach for a DSO.
In \cite{huang2016robust}, EV scheduling is studied under the worst-case scenario of wind generation uncertainty. 


In this paper, we present a coordinated framework for EV scheduling satisfying constraints of the grid. Indeed, our work extends the deterministic model  in \cite{original} by incorporating load and solar generation uncertainties. 
 While the integration of EVs has been widely explored, many of the existing studies focus on deterministic models or rely on single-method approaches for handling uncertainties, such as either SP or RO. However, these methods often fail to balance the trade-offs between computational efficiency and robustness, particularly when dealing with complex, real-world grid conditions that involve multiple sources of uncertainty. Our proposed framework, in contrast, introduces a hybrid SP-RO approach, which combines the strengths of both methods, allowing for more accurate and feasible EV scheduling that accounts for uncertainties in load demand and renewable generation.
 

 Thus, a hybrid RO/SP is applied which distinguishes our approach from the referenced works. In addition, this study accounts for different levels of uncertainty in the results to provide a more comprehensive analysis.  Our numerical results demonstrate that considering uncertainty significantly impacts the deliverable energy flexibility obtained from EV charging.

The rest of the paper is organized as follows. The system model is described in Section \ref{Model},  followed by the problem formulation in Section \ref{formu}. Section \ref{Simulation} presents the numerical results. Finally, the conclusions are discussed in Section \ref{Conclusion}.

\begin{table}[ht] 
\centering
\caption{NOTATIONS}
\begin{tabular}{|l|l|}
\hline
Notation   & Meaning\\
\hline
$k, n, t$ &  EV index, bus index, and time period index\\
\hline	
$d$, $g$, $l$   & Load index, generator index, and line index\\
\hline	
$\mathcal{T}, T$ & Set and number of time periods\\
\hline	
$\mathcal{K}, K$ & Set and number of EVs\\
\hline	
$\mathcal{K}_n$ & Set of EVs at bus $n$\\
\hline	
$\Phi, N$ & Set and number of buses\\
\hline	
$\Pi^{\sf L}_n$   & Set of lines connected to bus $n$\\
\hline	
$\Phi^{\sf G}_n$   & Set of generators located at bus $n$\\
\hline
$\xi^{\sf D}_{n}$   & Set of loads located at bus $n$\\
\hline	
$SOC_{n,k,t}$ & State of charge of EV $k$ at bus $n$ in period $t$\\ 
\hline
$\eta_{n,k}$ & Efficiency coefficient of EV $k$ at bus $n$\\
\hline
$P^{\sf EV}_{n,k,t}$ & Active power to charge EV $k$ at bus $n$ in period $t$\\
\hline
$Q^{\sf EV}_{n,k,t}$  & Injected reactive power for EV $k$ at bus $n$ in period $t$\\
\hline
$\alpha_t$ & Electricity price in period $t$ \\
\hline
$\Delta$ & Duration of one time period \\

\hline
$S_{n,k}^{EV}$ & Socket rating for EV $k$ at bus $n$  \\
\hline
$Sd_{n,k}$ & Desired SOC for EV $k$ at bus $n$ \\
\hline
$SOC^{\sf min}_{n,k}$ &Minimum  SOC for EV $k$ at bus $n$\\
\hline
$SOC^{\sf max}_{n,k}$ &Maximum  SOC for EV $k$ at bus $n$\\
\hline
$G_l,\beta_l$ & Conductance and susceptance of line $l$ \\

\hline
$V_{n,t}$ & Magnitude of voltage at bus $n$ in period $t$ \\
\hline
$\gamma_{n,t}$ & Voltage angle at bus $n$ in period $t$ \\
\hline
$P_{g,t},Q_{g,t}$ & Active/reactive power of generator $g$ at period $t$\\

\hline
$P_{l,t},Q_{l,t}$ & Active/reactive power flow at line $l$ in period $t$\\

\hline
$P_{d,t},Q_{d,t}$ & Active and reactive demand $d$ in period $t$\\

\hline
$P^{\sf gf}_{n,t}$ & Flexible active load at bus $n$ in period $t$ \\
\hline
$Q^{\sf gf}_{n,t}$ & Flexible reactive  load at bus $n$ in period $t$ \\
\hline
$P^{\sf pv}_{n,t}$ & Output active power of PV at bus $n$ at period $t$ \\
\hline
$V^{\sf min}_{n},V^{\sf max}_{n}$ & Minimum and maximum of voltage for bus $n$ \\
\hline
$S^{\sf max}_{g},S^{\sf max}_{l}$ & maximum capacity of generator and line  \\

\hline
\end{tabular} \label{notation}
\end{table}
\vspace{-0.75cm}
\section{System Model }
\label{Model}

We consider a power distribution network managed by a distribution grid control center.  There are also multiple EV aggregators located at different buses in the network. Each EV aggregator controls the charging schedule for a certain set of EVs. The grid control center communicates with the EV aggregators through a communication network. We consider $T$ time periods, where $\Delta$ is the duration of one time period. Let $t$ indicate the time period index. Also, $n, k$ are the bus index and EV index, respectively. Load, generator, and line indices are represented by $d, g,$ and $l$, respectively.

The set and the number of buses are $\Phi$ and $N$.  There is a set $\mathcal{K}$ of $K$ EVs in the system. 
Let $\mathcal{K}_n$ represent the set of EVs belonging to bus $n$. The set of lines, the set of generators, and the set of loads connected to bus $n$ are denoted by $\Pi^{\sf L}_n$, $\Phi^{\sf G}_n$, and $\xi^{\sf D}_{n}$, respectively.
$SOC_{n,k,t}$ indicates state of charge of EVs at each period, and $\eta_{n,k}$ is defined as charging efficiency coefficient for EV $k$. The active and reactive power of EV $k$ at bus $n$ at time $t$ are $P^{\sf EV}_{n,k,t}$ and $Q^{\sf EV}_{n,k,t}$.

The electricity price at time $t$ is denoted by $\alpha_t$, and $S^{\sf EV}_{n,k}$ is socket rating for EV $k$ at bus $n$.  Let $Sd_{n,k}$, $SOC^{\sf min}_{n,k}$, and $SOC^{\sf max}_{n,k}$ represent the desired state of charge (SOC), minimum and maximum SOC, respectively, for EV $k$ at bus $n$. The conductance and susceptance of line $l$ are $G_l$ and $B_l$. Voltage magnitude and its angle are represented by $V_{n,t}$ and $\gamma_{n,t}$, respectively. Also,
$V^{\sf min}_{n}$ and $V^{\sf max}_{n}$ represent the voltage restrictions.
Additionally, $P_{g,t},Q_{g,t}$, $P_{l,t},Q_{l,t}$, and $P_{d,t},Q_{d,t}$ are related to active and reactive power of generators, flow of lines, and demands, respectively. The flexible loads provided by the EV aggregator at bus $n$ are indicated by $P^{\sf gf}_{n,t}$ and $Q^{\sf gf}_{n,t}$. The solar PV energy  generation at bus $n$ at time $t$ is denoted by $P^{\sf pv}_{n,t}$. Finally, $S^{\sf max}_{g}$ and $S^{\sf max}_{l}$ indicate maximum generator and line capacity.

\begin{figure}[h!]
	\centering
		\includegraphics[width=0.4\textwidth,height=0.2\textheight]{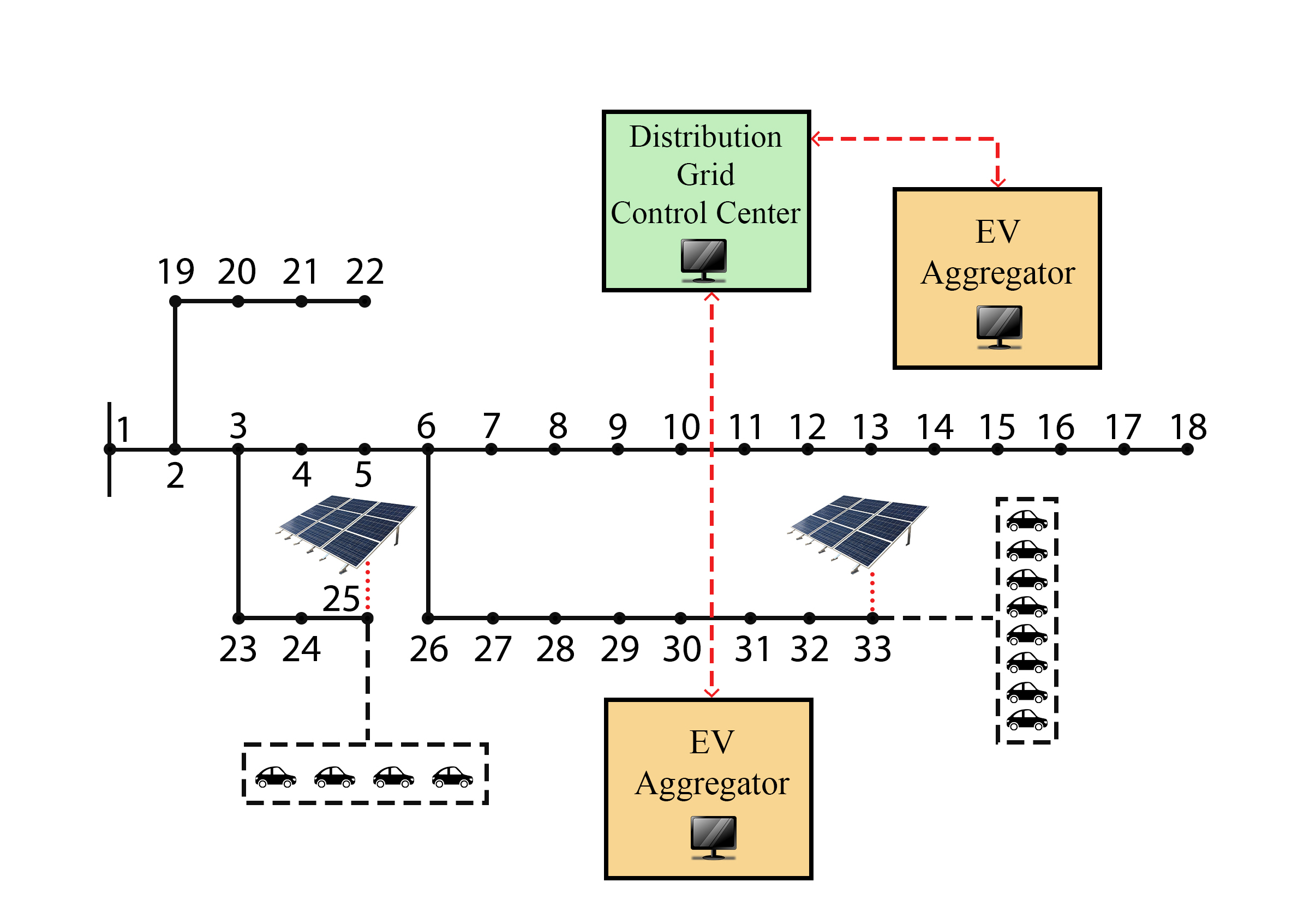}
			\caption{System Model}
	\label{fig:model}
\end{figure}

The proposed framework begins with the DSO  solving a grid model to optimize the deliverable energy flexibility achievable through EV aggregators. Subsequently, given the requested amount of flexible  EV charging loads computed by the DSO, each aggregator optimizes the EV charging for its fleet to minimize charging costs.
To support the grid effectively, EVs can operate in the first and fourth quadrants, injecting reactive power to increase allowable flexible loads without violating grid constraints, such as voltage drop and overloading.
Fig. \ref{fig:model} illustrates the coordinated schematic of the model, accounting for uncertainties in  demand and solar photovoltaic (PV) energy generation. 

\vspace{-0.45cm}

\vspace{-0.5cm}
\section{Problem Formulation}
\label{formu}
\vspace{-0.1cm}
\subsection{Distribution Grid Model}
\vspace{-0.0cm}

The DSO aims to solve the following optimization problem, with the objective of maximizing the flexible EV  load \cite{original}. 
\vspace{-0.0cm}
\beqn
\label{eq:dsoO}
\underset{P, Q, P^{\sf gf}}{\text{max}} \sum_t \sum_n P^{\sf gf}_{n,t}
\eeqn

subject to:
\begin{align}
&\label{eq:dsopl}
 P_{l,t} = \! G_l V^{\sf 2}_{l(s),t}\!-\! G_l V_{l(s),t} V_{l(r),t} \cos(\gamma_{l(s),t}-\gamma_{l(r),t}) \\ &
 + \beta_l V_{l(s),t} V_{l(r),t} \sin(\gamma_{l(s),t}-\gamma_{l(r),t}),~~~\forall l,t \nonumber \\&
 \label{eq:dsoql}
 Q_{l,t}  =\! \beta_l V^{\sf 2}_{l(s),t}-\beta_l V_{l(s),t} V_{l(r),t} \cos(\gamma_{l(s),t}-\gamma_{l(r),t}) \\ &
  - G_l V_{l(s),t} V_{l(r),t} \sin(\gamma_{l(s),t}-\gamma_{l(r),t}),~~~\forall l,t \nonumber \\ &
\label{eq:dsopb}
 \sum_{g\in \Phi^{\sf G}_n} \! \! P_{g,t}-\! \! \sum_{l\in \Pi^{\sf L}_n} \! \! P_{l,t} = \! \! P^{\sf gf}_{n,t}+\! \! \sum_{d\in \xi^{\sf d}_{n}} (P_{d,t}-P^{\sf pv}_{n,t}),~\forall n,t\\
&\label{eq:dsoqb}
\sum_{g\in \Phi^{\sf G}_n} Q_{g,t}-\sum_{l\in \Pi^{\sf L}_n} Q_{l,t} = Q^{\sf gf}_{n,t}+\sum_{d\in \xi^{\sf d}_{n}} Q_{d,t},~~~\forall n,t \\
&\label{eq:dsog}
P^{\sf 2}_{g,t}+Q^{\sf 2}_{g,t}\leq (S^{\sf max}_g)^{\sf 2},~~~\forall g,t \\
&\label{eq:dsol}
P^{\sf 2}_{l,t}+Q^{\sf 2}_{l,t}\leq (S^{\sf max}_l)^{\sf 2},~~~\forall l,t \\
&\label{eq:dsov}
V^{\sf min}_n\leq V_{n,t}\leq V^{\sf max}_n,~~~\forall n,t. 
\end{align}
In (\ref{eq:dsoO}), the flexible load should be maximized to accommodate more electric vehicles. This maximization will enhance the interaction between EVs and the grid, allowing for more effective voltage adjustment when facing voltage drop issues. Constraints (\ref{eq:dsopl}) and (\ref{eq:dsoql}) indicate equations regarding power transferred through the lines. Constraints (\ref{eq:dsopb}) and (\ref{eq:dsoqb})  represent the active and reactive power balance equations at each bus. 
Constraints (\ref{eq:dsog}) and (\ref{eq:dsol}) restrict the output power of generators and power flow through the lines, respectively. The limitations of bus voltages are captured by constraints (\ref{eq:dsov}). 

Unlike \cite{original} which assumes all system parameters are known, we consider the uncertainties of load and solar generation. Specifically, $P_{d,t}$ and $P_{n,t}^{\sf pv}$ are uncertain. In SP, a large number of scenarios can be generated to approximate the uncertainty. However, it may result in a large-scale problem that can be intractable. On the other hand, the RO approach can be quite conservative. 
To this end, we employ a hybrid stochastic/robust approach \cite{lin2004new}, \cite{janak2007new} to tackle the problem.
Additionally, the probability of failing to satisfy the constraints, which include uncertain parameters, must be less than the specified reliability level.

Specifically, the uncertainty-related constraints (\ref{eq:dsopb}) and (\ref{eq:dsoqb})  are equivalent to the following constraints with high probability \cite{lin2004new}, \cite{janak2007new}: 
\beqn
\label{eq:HSRP}
\sum_{g\in \Phi^{\sf G}_n} P_{g,t}-\sum_{l\in \Pi^{\sf L}_n} P_{l,t}\geq P^{\sf gf}_{n,t}+\sum_{d\in \xi^{\sf d}_{n}} (P_{d,t}-P^{\sf pv}_{n,t})\\ \nonumber - ~ \delta \max \{1,(|\sum_{d\in \xi^{\sf d}_{n}} (P_{d,t}-P^{\sf pv}_{n,t}))|\} \\ \nonumber+ \epsilon \lambda (\sum_{d\in \xi^{\sf d}_{n}} (P_{d,t}-P^{\sf pv}_{n,t})),\forall n,t\\
\label{eq:HSRQ}
\sum_{g\in \Phi^{\sf G}_n} Q_{g,t}-\sum_{l\in \Pi^{\sf L}_n} Q_{l,t} \geq Q^{\sf gf}_{n,t}+\sum_{d\in \xi^{\sf d}_{n}} Q_{d,t} \\ \nonumber 
- ~ \delta \max \{1,(|\sum_{d\in \xi^{\sf d}_{n}} Q_{d,t}|\}+\epsilon \lambda (\sum_{d\in \xi^{\sf d}_{n}} Q_{d,t}),~~~\forall n,t.
\eeqn

Due to space limitations,  we refrain from presenting the proof of the transformation here. However, this transformation is applicable when all uncertain parameters adhere to normal probability distribution functions (NPDF). 
Although electricity demands are commonly described by normal distributions, solar PV energy generation is typically modeled by a beta distribution.  To address this, we utilize the approximation technique from \cite{baharvandi2019linearized} to convert the beta distribution of solar generation into an NPDF.
Note that $\delta$ and  $\epsilon$ represent the infeasibility tolerance and uncertainty level, respectively, and the reliability level is determined by $\lambda$ of the uncertainty-related constraints. 
Replacing constraints (\ref{eq:dsopb}) and (\ref{eq:dsoqb}) with constraints (\ref{eq:HSRP}) and (\ref{eq:HSRQ}) transforms the DSO problem into a deterministic optimization problem. By solving this resulting problem, the DSO model provides optimal values of $P^{\sf gf}_{n,t}$ and $Q^{\sf gf}_{n,t}$, 
which are then communicated to the EV aggregators to optimize EV charging schedules.

\vspace{-0.5cm}
\subsection{Optimal Electric Vehicle Charging Model}
\vspace{-0.0cm}
Each EV aggregator $n$ at bus $n$ aims to minimize the charging cost of its EV fleet. 
The optimal EV charging problem for each aggregator can be expressed as follows:
\beqn
\zeta_n=\underset{P^{\sf EV}}{\text{min}} \sum_t  \sum_k \alpha_t P^{\sf EV}_{n,k,t} \Delta 
\eeqn

subject to:
\beqn
\label{ev:soc}
SOC_{n,k,t} = SOC_{n,k,t-1}+ \frac{\eta_{n,k} P^{\sf EV}_{n,k,t} \Delta}{E_{n,k}},~~~\forall n,k,t \\
\label{ev:pqs}
(P^{\sf EV}_{n,k,t})^{\sf 2}+(Q^{\sf EV}_{n,k,t})^{\sf 2}\leq 
(S^{\sf EV}_{n,k})^2,~~~\forall n,k,t \\
\label{ev:desired}
SOC_{n,k,t}\geq Sd_{n,k} ~~~,~~~\forall n,k,t = t_{n,k}^{\sf d}  \\
\label{ev:soclimits}
SOC^{\sf min}_{n,k}\leq SOC_{n,k,t} \leq SOC^{\sf max}_{n,k},~~~\forall n,k,t\\
\label{ev:PF}
\sum_k P^{\sf EV}_{n,k,t} \leq P^{\sf gf}_{n,t},~~~\forall n,t \\
\label{ev:QF}
\sum_k Q^{\sf EV}_{n,k,t} \geq Q^{\sf gf}_{n,t},~~~\forall n,t.
\eeqn
Note that the travel patterns, including the arrival and departure time for EVs, the desired SOC, and the initial and the final SOC are needed in the model. 
Constraints (\ref{ev:soc}) represent the energy dynamics of the EVs, where $E_{n,k}$ is the battery capacity of EV $k$ at bus $n$. 
Constraints (\ref{ev:pqs}) restrict the active and reactive power for each EV, considering its socket rating. 
Constraints (\ref{ev:desired}) enforce that the SOC level of each EV should be higher or equal to its desired SOC level at the time of departure $t_{n,k}^{\sf d}$.
To ensure good battery health, the battery SOC should be maintained within a certain allowable SOC range [$SOC_{n,k}^{\sf \min}, SOC_{n,k}^{\sf \max}$] as shown in (\ref{ev:soclimits}). 
Finally, constraints (\ref{ev:PF}) and (\ref{ev:QF}) set limits on the maximum total active power that EVs can consume and the maximum total reactive power that EVs can inject at bus $n$.
Here, $P_{n,t}^{\sf gf}$ and $Q_{n,t}^{\sf gf}$ are the solutions to the DSO model, which is sent to each EV aggregator for optimal EV charging. 

\vspace{-0.5cm}
\section{Simulation Results}
\label{Simulation}

In this work, we consider a time-slotted model with multiple 15-minute time periods. All the experiments are conducted in GAMS\footnote{{https://www.gams.com/}}. The DSO optimization problem is solved using the non-linear Knitro  solver\footnote{{https://www.gams.com/latest/docs/S\_KNITRO.html}} while the EV charging scheduling problem is solved using Mosek\footnote{{https://www.mosek.com/}}. To evaluate our model, we employ a modified IEEE 33-bus system, and the relevant data can be found in  \cite{soroudi2017power}. For generating arrival time, departure time, initial state of charge (SOC), and desired SOC of EVs, we utilize truncated Gaussian distribution functions with the following parameters:  $\mathcal{N}(0.66,0.1)$, $\mathcal{N}(0.82,0.1)$, $\mathcal{N}(48,26)$, and $\mathcal{N}(68,20)$, respectively \cite{data1}.
The Socket rating and charging efficiency for EVs are set to 11 kVA and 90\%, respectively. The maximum and minimum SOC are 80\% and 20\%. 
Details of electricity demand and prices can be seen in
\cite{soroudi2017power}. For simplicity, we consider only two EV aggregators located at buses 25 and 33.
The solar generation capacities of buses 25 and 33 are 0.15MW and 0.04MW, respectively \cite{Output}.  We assume there are 596 EVs at each of these two buses and the capacity of each EV is 30kWh.



\vspace{-0.3cm}
\begin{figure}[h!]
	\centering
		\includegraphics[width=3.5in, height=1.0in]{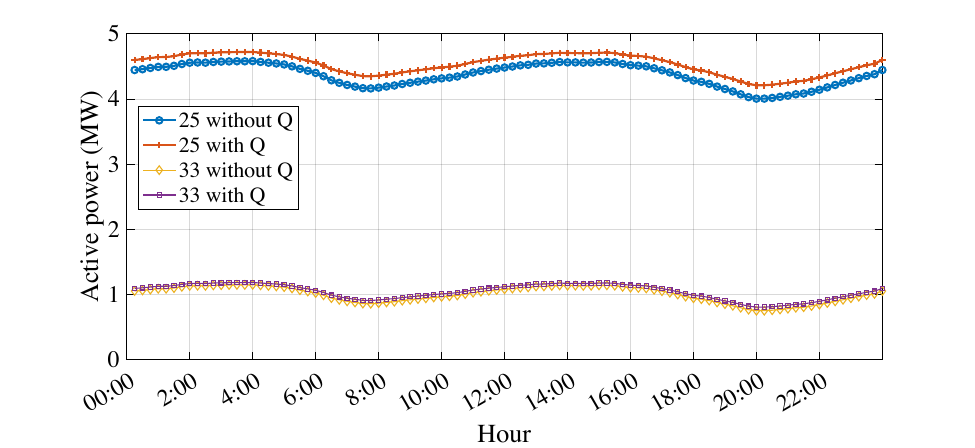}
  
			\caption{Allowable flexible active power at nodes 25 and 33 without uncertainty at unity and non-unity power factors}
	\label{fig:fig4}
\end{figure}

\vspace{-0.75cm}

\begin{figure}[h!]
	\centering 
		\includegraphics[width=3.5in, height=1.0in]{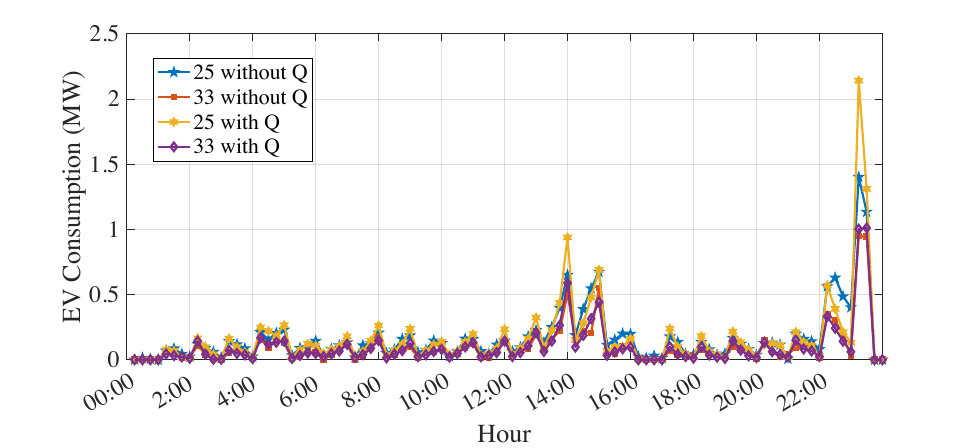}
			\caption{Consumed active power by EVs at nodes 25 and 33 with and without Q without uncertainty}
	\label{fig:fig5}
\end{figure} 
\vspace{-0.85cm}
\subsection{Grid Model and EV Scheduling Without Uncertainty}
In this case, the DSO solves the deterministic optimization problem (\ref{eq:dsoO})-(\ref{eq:dsov}) without considering uncertainties. We consider 
both unity and non-unity power factors (PF) for EVs. Note that unity PF implies the EVs do not inject reactive power into the grid. 
Fig. \ref{fig:fig4} depicts the optimal flexible EV charging 
loads at bus 25 and bus 33 over the scheduling horizon. 
We can observe that more EVs can be served in the case of non-unity PF. For instance, consider bus 25 in period 78, the active power that can be dedicated to EVs is  4.1184 MW for unity PF (i.e., without Q injected), while this amount is  4.2646 MW for non-unity PF (with Q injected). Similarly, consider bus 33 in period 78, 0.7767 MW can be delivered to EVs for unity PF, but it can increase to 0.8493 MW for non-unity PF. Additionally, the total flexible EV charging load (i.e., the optimal value of the objective function of the DSO)  changes from 518.6117 MW for EVs operated at unity PF to 535.4354 MW for non-unity PF. Fig. \ref{fig:fig5} represents the energy consumption of EVs at nodes 25 and 33 for unity and non-unity cases. For example, consider bus 33 in period 93, 0.947 MW can be absorbed by EVs at unity PF while this amount is 1 MW at non-unity PF.

\vspace{-0.35cm}

\subsection{Grid Model and EV Scheduling with Uncertainty}
\begin{figure}[http]
	\centering 
		\includegraphics[width=3.5in, height=1.0in]{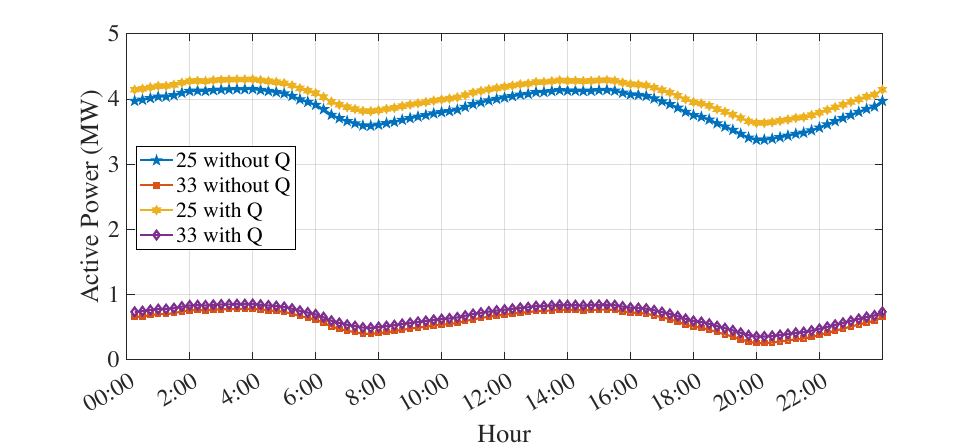}
			\caption{Allowable flexible active power at nodes 25 and 33 with uncertainty at unity and non-unity power factor}
	\label{fig:fig6}
\end{figure}

In this section, the uncertainty is taken into account by adding constraints (\ref{eq:HSRP}) and (\ref{eq:HSRQ}) in the DSO's problem.  We set the values of $\epsilon$, $\delta$, and $\lambda$ to 0.05, 0, and 6, respectively.  In Fig. \ref{fig:fig6}, the maximum active power provided for EVs is presented for nodes 25 and 33 in unity and non-unity PFs. For example, consider node 25 in period 76, the maximum active power is 3.572 MW for unity PF,  and it increases to 3.800 MW for non-unity PF, While the values are 4.199 MW and 4.336 MW, respectively, in the case without considering uncertainty, this highlights the significant impact that incorporating uncertainty has on the results. Similarly, the active power increases from 0.388 MW  to 0.482 MW for node 33, which confirms the advantage of operation at non-unity PF for EVs. 

\begin{figure}[h!]
	\centering 
		\includegraphics[width=3.5in, height=1.0in]{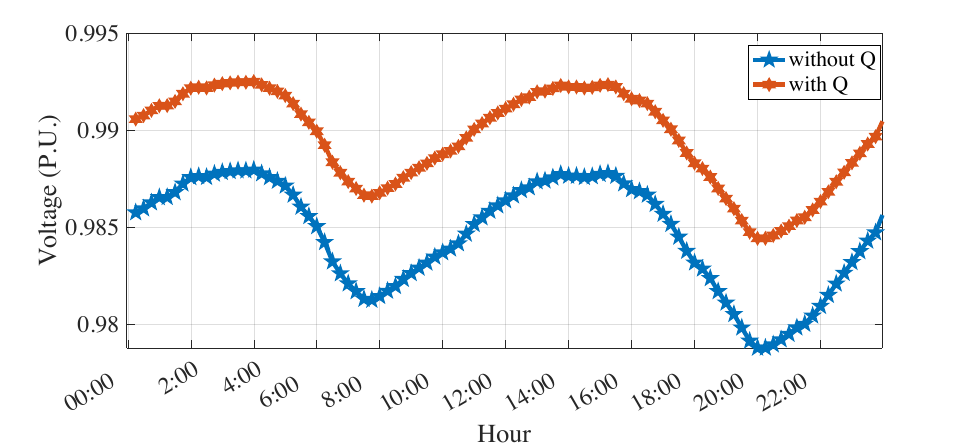}
			\caption{Voltage at node 17 with unity and non-unity power factor considering uncertainty}
	\label{fig:fig7}
\end{figure}

\begin{figure}[h!]
	\centering 
		\includegraphics[width=3.5in, height=1.0in]{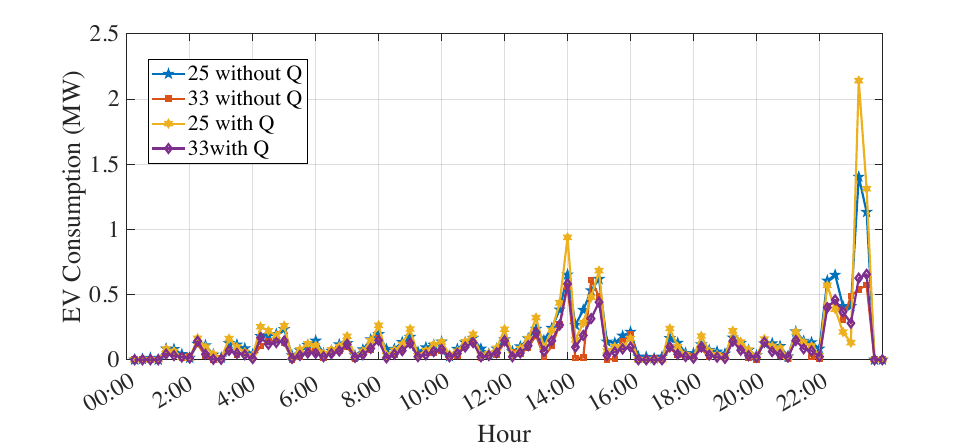}
			\caption{Consumed active power by EVs at nodes 25 and 33 with and without Q with uncertainty}
	\label{fig:fig8}
\end{figure}

Fig. \ref{fig:fig7} compares the level of voltage in bus 17 over the scheduling horizon with and without injecting reactive power, considering the system uncertainties. It illustrates the impact of injected reactive power by EVs on the voltage that would be able to help address voltage drops. For instance, the voltage in period 66 rises from 0.986 to  0.991 per unit when considering reactive power injection by EVs. 
Similar to the case without uncertainty, Fig. \ref{fig:fig8} represents the required active power for EVs at buses 25 and 33. For example, the consumed active power changes from 1.2 MW at unity PF in period 93 to 2.138 MW at non-unity PF for bus 25. 
EVs operating at a non-unity power factor can further enhance grid stability by providing reactive power support, which helps mitigate voltage drops.
\vspace{-0.5cm} 

\vspace{-0.5cm}
\vspace{-0.5cm}

\vspace{-0.2cm}
\begin{table}[h!] 
\centering
\caption{OPERATION COST}
\begin{tabular}{|l|l|l|l|l|}
\hline
Model  & WOUWOQ & WOUWQ & WUWOQ & WUWQ\\
\hline
Cost (\$) & 4640.744 & 4540.104 & 4660.217 & 4558.344\\
\hline

\end{tabular}
\label{TABII}
\end{table}
\vspace{-0.5cm}
\begin{figure}[h!]
	\centering 
	\includegraphics[width=3.5in, height=1.0in]{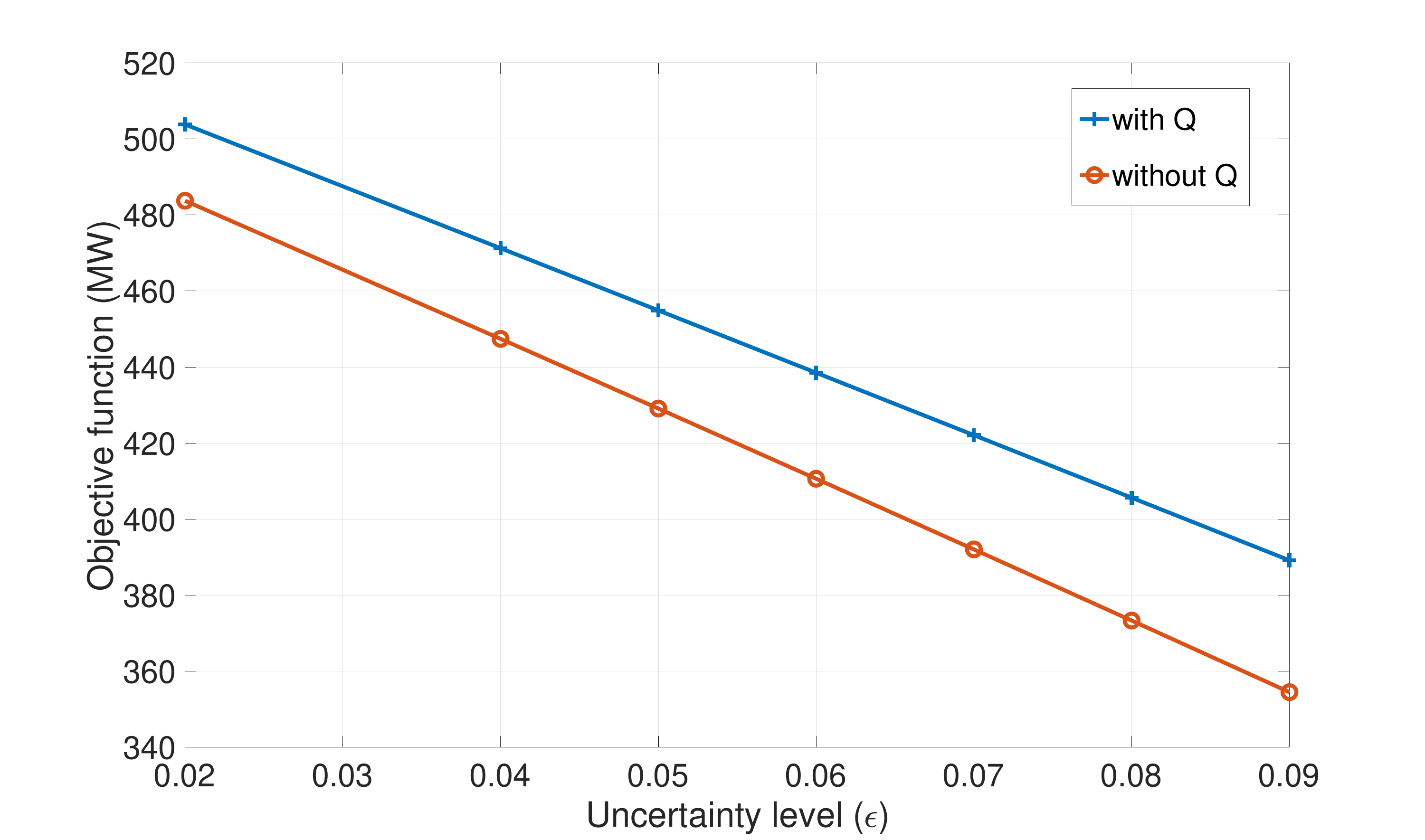}	
\caption{Effect of $\epsilon$ on allowable real power}
	\label{fig:fig9}
\end{figure} 
\vspace{0.1cm}
Table \ref{TABII} describes the operation costs of EVs without uncertainty without Q (WOUWOQ), without uncertainty with Q (WOUWQ), with uncertainty without Q (WUWOQ), and with uncertainty with Q (WUWQ) that by considering the uncertainty, the cost rises from \$4640 to \$4660 for the case without reactive power injection. 
From an economic perspective, the increase in operational costs when considering uncertainty reflects the real-world trade-offs between cost and reliability.
Finally, changes in $\epsilon$ can affect the allowable active power in the grid model. For instance, when reactive power is injected, increasing the level of uncertainty causes the objective function to decrease from 503 MW to 389 MW, as illustrated in Fig. \ref{fig:fig9}.

\vspace{-0.35cm}
\section{Conclusion}
\label{Conclusion}
In this paper, we proposed a two-level optimization framework where the DSO first solves the grid optimization problem to optimize the flexible loads for EV aggregators, considering the uncertainty of load and solar generation. Then, each aggregator aims to minimize the charging cost for its respective EV fleet. 
Since EVs would be operated at the fourth quadrant,  they can inject reactive power into the grid, which can help increase the allowable power dedicated to flexible loads. Furthermore, it can increase the voltage and prevent voltage drop. Due to the uncertain nature of PV and demand, a mixture of SP and RO has been applied to address the uncertainties. Our results show that the uncertainty leads to less allowable flexible loads. In our future work, we will consider the case where EVs can operate at the first quadrant to restrain over-voltage at buses.

\vspace{-0.1in}

\bibliographystyle{IEEEtran}
\bibliography{reference}

\end{document}